\documentstyle[12pt]{article}
\input epsf
\begin{document}
\hyphenation{ge-ne-ra-tes}
\hyphenation{me-di-um  as-su-ming pri-mi-ti-ve pe-ri-o-di-ci-ty}
\hyphenation{mul-ti-p-le-sca-t-te-ri-ng i-te-ra-ti-ng e-q-ua-ti-on}
\hyphenation{wa-ves di-men-si-o-nal ge-ne-ral the-o-ry sca-t-te-ri-ng}
\hyphenation{di-f-fe-r-ent tra-je-c-to-ries e-le-c-tro-ma-g-ne-tic pho-to-nic}
\title{
\hfill{\small IPNO/TH 94-38, May 1994}
\\ \hfill{\small (PRA-HEP 93/11, August 1993, extended)}
\vspace*{1.2cm}\\
\sc Density-of-states calculations and
multiple-scattering theory for photons
\vspace*{0.5cm}}
\author{\sc Alexander Moroz\thanks{
\tt Electronic address: moroz@fzu.cz, @ipncls.in2p3.fr}
\thanks{Present address:
School of Physics and Space Research, University of Birmingham,
Edgbaston, Birmingham B15 2TT, U. K.}
}
\date{
\protect\normalsize
\it Institute of Physics of the Academy of Sciences of the
Czech Republic,\\
\it Na Slovance 2, CZ-180 40 Praha 8, Czech Republic\\
 and\\
\it Division de Physique Th\'{e}orique\thanks{Unit\'{e} de Recherche des
Universit\'{e}s Paris XI et Paris VI
associ\'{e}e au CNRS}, Institut de Physique Nucl\'{e}aire,\\
\it Universit\'{e} Paris-Sud, F-91 406 Orsay Cedex, France
}
\maketitle

\addvspace{0.5cm}
\begin{center}
{\large\sc abstract}
\end{center}
The density of states for a finite or an infinite
cluster of scatterers in the case of both
electrons and photons can be represented in a general form
as the sum over all Krein-Friedel contributions of
individual scatterers  and a contribution due to the presence
of multiple scatterers. The latter is given by the sum over
all periodic orbits between different scatterers.
General three dimensional multiple-scattering theory for electromagnetic
waves in the presence of scatterers of arbitrary shape is presented.
Vector structure constants are calculated and general rules
for obtaining them from known scalar structure constants are given.
The analog of the Korringa-Kohn-Rostocker
equations for photons is explicitly written down.

\vspace*{0.3cm}

\vspace{0.2cm}

\begin{center}
{\em (Phys. Rev. B {\bf 51}, 2068 (1995))}
\end{center}
\thispagestyle{empty}
\baselineskip 20pt
\newpage
\section{Introduction}
\noindent
Multiple-scattering theory (MST) is a general method
for calculating the spectrum of either ordered or disordered
media and can deal with impurities as well \cite{PW}.
It is also rather physical since all that is needed to calculate
the spectrum in the presence of many identical scatterers is basically
scattering data ({\em phase shifts}) from a {\em single} scatterer
\cite{PW}. Moreover, the MST leads naturally to the
Korringa-Kohn-Rostocker (KKR) equations \cite{KKR}.
There are at least two important reasons that motivate
the  generalization of three-dimensional MST and KKR
to the case of {\em electromagnetic waves}. One of them is
motivated by the search for a {\em photonic band gap}
and related impurity problems in dielectric
lattices \cite{Y}. Low-lying photonic bands
have been calculated in two dimensions by some version of
the KKR method \cite{SM}, plane-wave method \cite{PSM},
and by the  transfer matrix method \cite{Pe}.
In three dimensions they have been quite efficiently calculated by
the plane-wave method and a large gap was shown to exist for the diamondlike
lattice \cite{HCS}.  Recently, photonic band gaps have been  predicted
numerically to exist for a body-centered-cubic $O_8$ lattice of cholesteric
blue phases characterized by tensor dielectric properties \cite{HSS1}.
An approximative Kohn-Luttinger method for photons was given
\cite{JH}. However, to calculate efficiently the density of states (DOS),
the impurity spectrum, and decide about the existence of a gap
even for the fcc lattice of dielectric spheres \cite{LL}
one needs more sophisticated methods like KKR \cite{SHI}.
In the latter example the first gap has been suggested to appear
between the eighth and the ninth
band that is quite in contradiction with the electronic case.
The second motivation for the three-dimensionsional  MST and KKR is the
{\em quantum chaos}  since the KKR method and its variations
are known to be numerically the most efficient method
to quantize various classically ergodic systems
such as quantum billiards \cite{MB,BGS}.

In the next section we start with the Lippmann-Schwinger integral
equations and derive the three-dimensional MST
for {\em electromagnetic waves}.
We shall consider a {\em non-conducting} medium with the current and
charge densities equal to zero.
Only the simplest isotropic case is considered, ${\bf D}({\bf r}) =
\epsilon({\bf r}){\bf E}({\bf r})$ and
${\bf B}({\bf r})=\mu({\bf r}){\bf H}({\bf r})$ where
${\bf D}({\bf r})$ (${\bf B}({\bf r})$) is the electric (magnetic)
induction. We shall confine ourselves to monochromatic waves
to avoid the nonlocal time relation between ${\bf D}({\bf r})$ and
${\bf E}({\bf r})$,
or ${\bf B}({\bf r})$ and ${\bf H}({\bf r})$. We shall also
allow for {\em complex}  permeabilities, i.\,e., for an absorption.
In such medium the {\em stationary macroscopic} Maxwell equations
are  symmetric under \cite{BW}
\begin{equation}
{\bf E}({\bf r})\rightarrow{\bf H}({\bf r}),
\hspace{0.5cm}{\bf H}({\bf r})\rightarrow
-{\bf E}({\bf r}),\hspace{0.5cm}
\epsilon({\bf r})\rightarrow\mu({\bf r}).
\label{symm}
\end{equation}
We shall not fix the host permittivity
$\epsilon_o=1$ in order to be also able to consider
voids. For simplicity magnetic permeability $\mu({\bf r})$
will be set to $1$.
Then the Maxwell equation for the electric intensity
${\bf E}$ can be written as
\begin{equation}
\mbox{\boldmath $\nabla$}\times[\mbox{\boldmath $\nabla$}\times
{\bf E}({\bf r})]
-(\omega/c)^2 v({\bf r})
{\bf E}({\bf r})= (\omega/c)^2\epsilon_o\, {\bf E}({\bf r}),
\label{ME*}
\end{equation}
where $v({\bf r})=\epsilon({\bf r})-\epsilon_o$ is an analog of
a {\em potential}. For notational simplicity we shall set $c=1$
from now on.
 Scatterers are allowed to have {\em arbitrary} shape and to be
{\em arbitrarily} arranged in a host homogeneous dielectric medium
$\Omega$ with permittivity $\epsilon_o$
up to the usual constraint of MST : it will always be assumed
that $\Omega$ can be divided into {\em nonoverlapping} spheres
$V_n{}'^s$ each of which contains {\em one and only one} scatterer.
The number $N$ of scatterers may be either {\em finite} or {\em infinite}.
The generalization to general non-muffin-tin potentials
(i.e., overlapping spheres) can be
performed following Faulkner \cite{JSF}.
%

In contrast to electrons,
in the case of photons one can almost safely ignore photon-photon
interaction and rather full control of the scattering
potential $v({\bf r})=\epsilon({\bf r})-\epsilon_o$ can be achieved
in experiment.  As a result
the single photon approximation, the photonic band structure, as well
as the standard assumptions of the MST turn out to be {\em exact}.
Therefore, experiments on photons on well-controlled samples
are appealing for both, testifying the basic ideas about
the semiclassical quantization \cite{MG} and
to provide the check on the MST itself.
The KKR equations and structure constants are explicitly written
down. A relation with the standard electronic structure constants
is maintained as much as possible. General rules
are formulated as to how the photonic structure constants can be obtained
from the electronic ones.

In Sec. \ref{oper} the operator content of the MST is discussed in
more  detail and  the connection
with standard calculations within
the context of the Schr\"{o}dinger Hamiltonian is made.
Finally, in Sec. \ref{path} methods are
presented for the calculation of the DOS for an ensemble
of scatterers. The results are applicable for both electrons and photons.
A more detailed analysis of the Lloyd and Smith
on-the-energy-shell formalism \cite{PW} from the point of view of
the Gutzwiller trace formula \cite{MG} is made.
Surprisingly enough it is found  that similar to the latter case,
the DOS can be calculated by summing over all {\em `periodic
orbits'} in the `phase space'. One speaks about the
`phase space' here in a symbolic sense which arises
as the consequence of the on-the-energy-shell formalism.
The `periodic orbits', which start and end with the
same coordinate, angular momentum, and multipole indices,
are made from all possible
connections ({\em random walks}) between the centers of different scatterers
with a possible return.

The Krein-Friedel formula \cite{PW,F} is discussed, too.
Recently we have shown that when it is supplemented with
$\zeta$-function regularization it can be used for some infinite
range potentials in the {\em singular} scattering problem like the
Aharonov-Bohm potential
\cite{AM1}. Here we shall show that it can be used for
the Maxwell equations as well.

\section{Multiple-scattering theory and the KKR  equations for photons}
\label{mstsec}
The Maxwell equation (\ref{ME*}) for ${\bf E}$ resembles
the Schr\"{o}dinger equation \cite{Y}. However, the behavior
of ${\bf E}$ across a discontinuity $\Sigma$ of $\epsilon$ is essentially
different. Instead of the {\em continuity} of fields and their
derivatives one has
a {\em discontinuity} of $E_n({\bf r})$, the normal component
of ${\bf E}$, and the continuity of
$E_t({\bf r})$, the tangential component of ${\bf E}({\bf r})$, on
$\Sigma$ \cite{BW},
\begin{equation}
E_n^+ - E_n^-=\frac{\epsilon_- -\epsilon_+}{\epsilon_+} E_n^-
=\frac{v({\bf r})}{\epsilon_o}E_n^-, \hspace{0.8cm}E_t^+=E_t^-.
\label{uno}
\end{equation}
The derivative $\mbox{\boldmath $\nabla$}_t E_t$ is continuous,
$\partial_n {\bf E}_t$ and $\mbox{\boldmath $\nabla$}_t E_n$
are {\em discontinuous} \cite{AM},
\begin{equation}
\left.(\partial_n {\bf E}_t^+ -\partial_n {\bf E}_t^-)\right|_\Sigma
=\left(\frac{1}{\epsilon_+}-
\frac{1}{\epsilon_-}\right) \mbox{\boldmath $\nabla$}_t D_n
= \frac{v}{\epsilon_o}\mbox{\boldmath $\nabla$}_t E_n^-
=\left.(\mbox{\boldmath $\nabla$}_t E_n^+  -
\mbox{\boldmath $\nabla$}_t E^-_n)\right|_\Sigma,
\end{equation}
while $\partial_n {\bf E}_n({\bf r})$  obeys \cite{AM}
\begin{equation}
\left.(\partial_nE_n^+ -\partial_nE_n^-)\right|_\Sigma =-
\left(\frac{1}{\epsilon_+^2}(\partial_n\epsilon)_+ -\frac{1}
{\epsilon_-^2}(\partial_n\epsilon)_-\right)D_n.
\end{equation}
Thus,  although the normal component $E_n({\bf r})$ of ${\bf E}({\bf r})$ is
discontinuous,
the side limits of $\partial_nE_n({\bf r})$ {\em coincide} provided the side
limits of the derivatives of $\epsilon({\bf r})$ at $\Sigma$ are zero, as
is the case of the {\em muffin-tin} (piecewise constant)
potential. The above relation have been shown to be
 a general consequence of
the (nonstationary) Maxwell equations in a dielectric \cite{AM}.

That, generically {\em discontinuous} behavior of ${\bf E}$ and its
derivatives, necessitates the introduction of the concept of {\em outward}
and {\em inward integral equations} \cite{AM}.
One has to distinguish  strictly  between them.
The inward formalism is basically that of  Kohn-Korringa-Rostocker
(KKR) (Ref. \cite{KKR})
while the outward formalism is originally due to Morse \cite{M}.
In the case of the Schr\"{o}dinger equation both formalisms coincide.

One way to derive the MST there is to use {\em outward} integral
equations  under the presence of an incident wave ${\bf E}_o({\bf r})$.
To keep track with the standard derivation \cite{PW} one
starts with the Lippmann-Schwinger equation for ${\bf E}({\bf r})$ \cite{AM},
\begin{eqnarray}
\lefteqn{{\bf E}({\bf r})= {\bf E}_o({\bf r}) -\omega^2\,\sum_n
\left\{\int_{V_n} G_{\sigma}({\bf r} ,{\bf r} ')\,v({\bf r}')\,
{\bf E}({\bf r} ')\,d^3{\bf r} '-
\right.}
\hspace{2.5cm}\nonumber\\
&&\left.\mbox{} -\int_{V_{n+}} [\mbox{\boldmath $\nabla$}'
G_{\sigma}({\bf r} ,{\bf r} ')] [\mbox{\boldmath $\nabla$}'
\cdot{\bf E} ({\bf r} ')] \,d^3{\bf r} '\right\},
\label{entq1}
\end{eqnarray}
with $n$ labeling scatterers. $V_{n+}$ is a shorthand for the outward
limit through measurable sets $\Sigma_\ell$,
$\Omega\supset\Sigma_\ell\supset V_n$, $\lim_{\ell\rightarrow\infty}
\Sigma_\ell\searrow V_n$. Here $G_{\sigma}({\bf r} ,{\bf r} ')$ is
the Green function of the Helmholtz equation,
\begin{equation}
(\mbox{\boldmath $\nabla$}^2 +\sigma^2)G_\sigma ({\bf r} ,{\bf r} ')
= \delta ({\bf r} -{\bf r} '),
\label{green}
\end{equation}
$\sigma^2$ being $\sigma^2=\omega^2\epsilon_o$,  $\omega$ is
a given frequency of a photon.
$G_\sigma ({\bf r} ,{\bf r} ')$ is chosen to satisfy the
scattering conditions,
\begin{equation}
G_\sigma ({\bf r} ,{\bf r} ') = G^+_\sigma ({\bf r} ,{\bf r} ')
=-\frac{1}{4\pi}\frac{e^{i\sigma|{\bf r}
-{\bf r}'|}}{|{\bf r}-{\bf r}'|}\cdot
\label{greenf}
\end{equation}
Equation (\ref{entq1}) is essentially the same
(up to an integration per parts)  as Eq.\ (4) in Ref. \cite{XZ}
written in terms of the tensorial Green function
${\bf d}_o({\bf r},{\bf r}')$,
\begin{equation}
{\bf d}_o({\bf r},{\bf r}')=-\left[\omega^2{\bf 1}
+\frac{1}{\epsilon_o}\mbox{\boldmath $\nabla$}
\otimes\mbox{\boldmath $\nabla$}\right]G_\sigma ({\bf r} ,{\bf r} '),
\label{d0}
\end{equation}
where $\mbox{\bf 1}$ and $\mbox{\boldmath $\nabla$}\otimes\mbox{\boldmath
$\nabla$}$ are, respectively, $3\times 3$ identity
and $(\partial_i\partial_j)$ matrices.
When the volume integrals are  rewritten in terms of the
surface integrals and provided ${\bf r}$ stays {\em inside}
the spheres one finds \cite{AM}
\begin{equation}
{\bf E}_o({\bf r})=
\sum_n\oint_{\partial V_{n+}}dS' [\partial_{r '}
G_\sigma ({\bf r},{\bf r} ') -
G_\sigma({\bf r}, {\bf r} ')\,\partial_{r '}]\, {\bf E} ({\bf r} '),
\label{other1}
\end{equation}
which is nothing but the $on-shell$ Lippmann-Schwinger equation
(see also Sec. \ref{oper}).
It is probably here when we differ from Ref. \cite{XZ}. It is, however,
difficult to say definitely since, in \cite{XZ},
 they do not indicate on which
side of the boundary their surface integrals are.

The basic idea of MST
(and the KKR method as well) is to rewrite the {\em integral} equation
into an {\em algebraic} one. This is accomplished by expanding
the Green function and electric field in the basis of spherical harmonics.
To make the integration in (\ref{other1}) well defined we have to
choose some ${\bf r}$ and divide $\Omega$ into regions where,
respectively, $r>r'$ or $r<r'$.
The reason is that the expansion of the Green function
$G_\sigma ({\bf r} ,{\bf r} ')$  into spherical harmonics
does depend as to whether $r<r'$ or $r>r'$,
\begin{eqnarray}
G_\sigma ({\bf r} ,{\bf r} ') =-i\sigma\sum_{L}[h^+_l(\sigma r)j_l(\sigma
r')\Theta(r-r')
+j_l(\sigma r)h^+_l(\sigma r')\Theta(r'-r)]
\,\nonumber\\
\mbox{} \times
Y_{L}(\theta,\phi)Y^*_{L}(\theta',\phi').
\label{greenexp}
\end{eqnarray}
Here $\Theta(x)$ is the usual Heaviside step function, $L$ is actually
multi-index $L=lm$, and $j_l$ and $h^+_l=j_l +in_l$ are respectively
the spherical Bessel and the spherical Hankel function, $n_l$
being the spherical Neumann function \cite{PW,J}.
Since
\begin{equation}
\sum_{m=-l}^l Y_{L}(\theta,\phi)Y^*_{L}(\theta',\phi')=
\sum_{m=-l}^l Y^*_{L}(\theta,\phi)Y_{L}(\theta',\phi')
\end{equation}
the arguments of spherical harmonics can be interchanged in
Eq. (\ref{greenexp}).
In what follows we shall denote by ${\bf R}_n$ the center of
the $n$th sphere $V_n$, and by $r_n$ its radius.
Without  any restriction ${\bf r}$ can be assumed to be such that
$\forall\, n\neq 0$: $|{\bf r}-{\bf R}_o|<r_o<|{\bf r}-{\bf R}_n|$.
Due to translational invariance of the free Green function (\ref{greenf})
${\bf R}_o$ can be set to  origin. Therefore,  on the surface of all spheres
the expansion (\ref{greenexp}) with $r'>r$ will be used in (\ref{other1}).

The use of the Lippmann-Schwinger equation requires that the incident wave
be {\em singularity free} in the whole space \cite{PW,RN}.
Therefore, ${\bf E}_o({\bf r})$ is assumed to have an expansion of the form
\begin{equation}
|{\bf E}_o({\bf r})\rangle=
\sum_{AL}C^{o}_{oAL} |{\bf J}_{AL}({\bf r})\rangle ,
\label{inexp}
\end{equation}
with $C^{o}_{oAL}$ expansion constants
and the parameter $A$ labeling electric and magnetic multipoles,
\begin{equation}
|{\bf J}_{ML}({\bf r})\rangle = j_{l}(\sigma r)|{\bf Y}^{(m)}_L\rangle ,
\label{mmulti}
\end{equation}
\begin{eqnarray}
\lefteqn{|{\bf J}_{EL}({\bf r})\rangle = \frac{i}{(2l+1)\sqrt{\epsilon_o}}
\left\{j_{l-1}
(\sigma r)
\left[\sqrt{l(l+1)}|{\bf Y}^{(o)}_{L}\rangle+ (l+1)|{\bf Y}^{(e)}_{L}\rangle
\right]\right.}\hspace{2cm}\nonumber\\
&&
\left.\mbox{} + j_{l+1}(\sigma r)\left[
\sqrt{l(l+1)}|{\bf Y}^{(o)}_{L}\rangle-
l|{\bf Y}^{(e)}_{L}\rangle\right]\right\}.
\label{elmd}
\end{eqnarray}
Here  $|{\bf Y}^{(a)}_{L}\rangle$ are  normalized {\em magnetic} ($a=m$),
{\em electric} ($a=e$), and
{\em longitudinal} ($a=o$) vector spherical harmonics (see Appendix
\ref{nota}). Note that there are neither magnetic nor electric multipoles
for $l=0$.
The above form of the electric multipole is `canonical' one since
in the case of constant $\epsilon(r)$ and $\mu(r)$
the Maxwell equations determine a relation
between scalars ${\bf r}\cdot{\bf H}$ and
${\bf L}\cdot{\bf E}$, and ${\bf r}\cdot{\bf E}$ and ${\bf L}\cdot{\bf H}$
\cite{J},
\begin{equation}
{\bf r}\cdot{\bf H} = \frac{1}{\omega\mu}\,
{\bf L}\cdot{\bf E} ,\hspace{2cm}
{\bf r}\cdot{\bf E} = -\frac{1}{\omega\epsilon}\,{\bf L}\cdot{\bf H}.
\end{equation}
Here, ${\bf L}$ is the orbital angular momentum operator,
${\bf L}=-i{\bf r}\times\mbox{\boldmath{$\nabla$}}$.
In what follows, however, the factor $\sqrt{\epsilon_o}$
in (\ref{elmd}) will be rescaled for it will simplify the
resulting formulas.

Now, $|{\bf E}({\bf r})\rangle$ is expanded near a scatterer centered at
${\bf R}_o$ in the basis of final states $|{\bf F}_{AL}({\bf r})\rangle$,
\begin{equation}
|{\bf E}({\bf r})\rangle=\sum_{AL}C_{oAL} |{\bf F}_{AL}({\bf r})\rangle .
\label{finexp}
\end{equation}
Due to (\ref{other1}) one needs
to expand $|{\bf E}({\bf r})\rangle$ only for ${\bf r}$ on the boundaries of
spheres. For a given $n$th sphere one has
\begin{equation}
|{\bf E}({\bf r})\rangle=\sum_{AL} C_{nAL} |
{\bf F}_{nAL}({\bf r}-{\bf R}_n)\rangle ,
\label{rfinexp}
\end{equation}
where
\begin{equation}
|{\bf F}_{nAL}({\bf r})\rangle = |{\bf J}_{nAL}({\bf r})\rangle -
\oint_{\partial V_{n+}}dS' [\partial_{r '} G_\sigma ({\bf r},{\bf r} ') -
G_\sigma({\bf r}, {\bf r} ')\,\partial_{r '}]|{\bf F}_{nAL}({\bf r}')
\rangle.
\end{equation}
On the {\em outward} side of the scatterer then
\begin{equation}
|{\bf F}_{nAL}({\bf r}_n)\rangle = |{\bf J}_{AL}({\bf r}_n)\rangle -i
\sum_{A'L'}t^n_{A'L',AL}|{\bf H}^+_{A'L'}({\bf r}_n)\rangle,
\end{equation}
where $t^n_{A'L',AL}$ is the  {\em transition} or $t$-{\em matrix} of
the $n$th scatterer \cite{PW}, and ${\bf r}_n ={\bf r}-{\bf R}_n$ is
the radius vector on the boundary of the sphere $V_n$.

To find a more convenient expansion of $G_\sigma({\bf r}, {\bf r} ')$
on the boundary of a given $n$th sphere, $n\neq 0$,
one can use the expansion of the (scalar) {\em scattered} wave from
the scatterer centered at ${\bf R}_o=0$ into an {\em incident} wave
about the scatterer centered at ${\bf R}_n$,
\begin{equation}
h^+_l(\sigma r)Y^*_L({\bf r}) =
\sum_{L''}j_{l''}(\sigma|{\bf r}-{\bf R}_n|)Y^*_{L''}({\bf r}-{\bf R}_n)
\,g_{L''L}({\bf R}_n),
\label{ino}
\end{equation}
with $g_{L''L}({\bf R}_n)$ being the {\em scalar structure constant}
\cite{PW}, the very same as in the case of electrons. They are given
by
\begin{equation}
g_{L'L}({\bf R})
 = - (-1)^{l'} i^{l+l'+1}  4\pi \sigma\sum_{L_1} C_{L'L}^{L_1}(-i)^{l_1}
 h^+_{l_1}(\sigma R) Y^*_{L_1}({\bf R}).
\label{stcon}
\end{equation}
The numerical constants $C_{LL'}^{L_1}$  here are the {\em Gaunt numbers}
\cite{PW}. For a given pair of ${\bf r}$ and ${\bf r}'$ such that
$|{\bf r} -{\bf R}_o|<|{\bf r} -{\bf R}_j|$ $\forall j\neq 0$, and
$|{\bf r}' -{\bf R}_1|<|{\bf r}' -{\bf R}_j|$ $\forall j\neq 1$,
the expansion (\ref{greenexp}) of the Green function
$G_\sigma ({\bf r},{\bf r}')$ can be rewritten as follows,
\begin{equation}
G_\sigma ({\bf r},{\bf r}') = \sum_{L,L'}
g_{L'L}({\bf R}_1 -{\bf R}_o) j_l(\sigma r) Y_L({\bf r})
j_{l'}(\sigma |{\bf r}' -{\bf R}_1|) Y^*_{L'}({\bf r}' -{\bf R}_1) .
\label{grstrex}
\end{equation}

Now (\ref{inexp}), (\ref{finexp}), (\ref{grstrex}) and
$dS'=r_n^2d\Omega$, $r_n$ being the radius
of the $n$th sphere, are inserted in
(\ref{other1}). After some manipulations one finds
\begin{eqnarray}
\lefteqn{\sum_{AL}C^o_{oAL}|{\bf J}_{AL}({\bf r})\rangle =
i\sigma  \sum_{ALL'}j_l(\sigma r)Y_L\,
r_o^2\,\langle Y_{L}|W[h^+_{l},
{\bf F}_{oAL'}]\rangle_o\, C_{oAL'} }\nonumber\\
&&
\mbox{}+
i\sigma
\sum_{n\neq 0,ALL'L''}j_l(\sigma r)Y_L\,
r^2_n\, \langle Y_{L''}|W[j_{l''},
{\bf F}_{nAL'}]\rangle_n\, g_{L''L}({\bf R}_n)\,C_{nAL'}.\hspace{0.3cm}
\label{big1}
\end{eqnarray}
Here $\langle|\, .\,| \rangle_n$ denotes integration over
angle variables on the $n$th sphere.
Index $n$ means that the value of spherical functions is taken
at $r_n$, the radius of the $n$th sphere.  $W(u,w)$ is the {\em Wronskian},
\begin{equation}
W(u,w)=(u\,\partial_r -\partial_r u)\,w = u\,\partial_r w- \partial_r u\,w.
\end{equation}

By taking the scalar product, after approaching
$r_o$ by $r$ from below, and by using identities from
Appendix \ref{basic} one can rewrite (\ref{big1}) as
\begin{equation}
C^o_{oAL} N_A^l = \sum_{jBKA'L'}\left[
\delta_{oj}\delta_{AA'}\delta_{LL'}N_A^l
- \bar{G}^{oj}_{AL,BK}t^j_{BK,A'L'}\right]C_{jA'L'},
\label{ito}
\end{equation}
where
\begin{equation}
N_A^l=\left\{
\begin{array}{cc}
j_l^2(\sigma r_o), &A=M\\
\frac{1}{(2l+1)}\left[(l+1)j_{l-1}^2(\sigma r_o)
+ lj_{l+1}^2(\sigma r_o)\right],  &A=E.
\end{array}
\right.
\label{na}
\end{equation}
Here the quantity $\bar{G}^{ij}_{AL,BK}$ has been introduced,
\begin{eqnarray}
\lefteqn{\bar{G}^{ij}_{AL,BK}=\bar{G}_{AL,BK}({\bf R}_j-{\bf R}_i) =
i\sigma r_j^2\sum_{L'}\sum_{L"} j_{l'}(\sigma r_i)}\nonumber\\
&&
\mbox{}\times
\langle{\bf J}_{AL}|Y_{L'}\rangle_i\cdot\langle Y_{L"}|
W[j_{l''},{\bf H}^+_{BK}]\rangle_j\, g_{L"L'}({\bf R}_j -{\bf R}_i).
\label{vectorgreen}
\end{eqnarray}
To calculate
$\bar{G}^{ij}_{AL,A'L'}$ explicitly we shall use the basic
MST identities collected in Appendix \ref{basic}.
By defining quantities $\bar{C}^\alpha$ in terms of $3j$ symbols,
\begin{eqnarray}
\bar{C}^\alpha(l,-1,m)&=&\sqrt{l+1}\,C^\alpha(l-1,m+\alpha,l,m),
\nonumber\\
\bar{C}^\alpha(l,1,m) &=& -\sqrt{l}\,C^\alpha(l+1,m+\alpha,l,m),
\end{eqnarray}
and
\begin{equation}
\bar{T}^\alpha_{lm}=\frac{1}{\sqrt{l(l+1)}}\,T^\alpha_{lm}
\end{equation}
one can write
the resulting expressions in the following compact form,
\[
\bar{G}^{ij}_{ML,ML'}= j_l^2(\sigma r_i)\sum_{\alpha=-1}^1
g^{ij}_{l'm'+\alpha;lm+\alpha}\,
\,\bar{T}^\alpha_{l'm'}\,\bar{T}^\alpha_{lm},
\]
\[
\bar{G}^{ij}_{ML,EL'}=
j_l^2(\sigma r_i) \sum_{\stackrel{p'=-1}
{\scriptscriptstyle p'\neq 0}}^1 \sum_{\alpha=-1}^1
g^{ij}_{l'+p',m'+\alpha,lm+\alpha}
\bar{C}^\alpha(l',p',m')\,
\bar{T}^\alpha_{lm},
\]
\[
\bar{G}^{ij}_{EL,ML'}=
\sum_{\stackrel{p=-1}
{\scriptscriptstyle p\neq 0}}^1 j_{l+p}^2(\sigma r_i) \sum_{\alpha=-1}^1
g^{ij}_{l',m'+\alpha,l+p,m+\alpha}
\bar{T}^\alpha_{l'm'}\,
\bar{C}^\alpha(l,p,m),
\]
\begin{eqnarray}
\lefteqn{
\bar{G}^{ij}_{EL,EL'}= \sum_{\stackrel{p=-1}
{\scriptscriptstyle p\neq 0}}^1 j_{l+p}^2(\sigma r_i)
\sum_{\stackrel{p'=-1}{\scriptscriptstyle p'\neq 0}}^1
 \sum_{\alpha=-1}^1
g^{ij}_{l'+p',m'+\alpha,l+p,m+\alpha}
}
\hspace{2cm}\nonumber\\
&&\mbox{}\times
\bar{C}^\alpha(l',p',m')\,\bar{C}^\alpha(l,p,m).
\label{structure}
\end{eqnarray}
The quantity $\bar{G}^{ij}_{AL,BK}$ is not yet the vector structure
constant $G^{ij}_{AL,BK}$ we are looking for. It depends on
the muffin-tin radius $r_i$ that enters the argument of
Bessel functions. However, one sees immediately, by comparing
(\ref{ito}) and (\ref{na}) with (\ref{structure}), that $N_M^l$ can be
rescaled. To rescale $N_E^l$ is more tricky.
By using the relations 10.1.19, 10.1.21, 10.1.22 of Ref. \cite{AS} one can
show that
\begin{equation}
N_E^l(z)= \left(\frac{j_l(z)}{z}
+\sigma j_l'(z)\right)^2
+l(l+1)\frac{j_l^2(z)}{z^2},
\label{nel}
\end{equation}
\begin{equation}
j_{l-1}^2(z) = N_E^l(z) +l\,\frac{j_l(z)}{z}
\left(\frac{j_l(z)}{z}+2\sigma j_l'(z)\right),
\label{term1}
\end{equation}
\begin{equation}
j_{l+1}^2(z)= N_E^l(z) -(l+1)
\frac{j_l(z)}{z}
\left(\frac{j_l(z)}{z}+2\sigma j_l'(z)\right),
\label{term2}
\end{equation}
where $z=\sigma r_o$, and prime means the derivative with respect to $r$.
According to (\ref{nel}) $N_E^l(z)$ is the sum of two non-negative terms.
By using the relation 9.5.2. of Ref. \cite{AS} about the interlace of zeros
of $j_l$ and $j_l'$ one can show that, provided $z\neq 0$, always $N_E^l>0$.
Indeed, if first term is zero then the second is nonzero and {\em vice versa}.
The latter two equations give the expression
of the relevant Bessel functions that enter (\ref{structure})
in term of $N_E^l$. When columns of
$\bar{G}^{ij}_{AL,BK}$ with $E$ labels are divided by $N_E^l$
one has to show that the contributions of the second terms in
(\ref{term1}) and (\ref{term2}) cancel. By close inspection one finds
that they introduce respectively the factor $l$ and $-(l+1)$
to (\ref{structure}). By using the definition (\ref{stcon}) of the scalar
structure constant $g_{L'L}({\bf R})$ and the properties of $3j$ symbols
one finds that the contributions indeed cancel.
The rescaled structure constants will be written without bar.
Physically $G^{ij}_{AL,A'L'}$ describes what amount of a particular
multipole field
{\em scattered} from the $i$th site contributes to the particular
multipole field {\em incident} on the $j$th site.
Relations  (\ref{structure}) imply the following rules for obtaining
$G^{ij}_{AL,A'L'}$ from scalar $g^{ij}_{L'L}{}'^s$ :

\vspace*{0.3cm}

{\bf 1.} {\em If $A=M$ then replace $L$ in $g^{ij}_{L'L}$ by $L=l,m+\alpha$
and multiply $g^{ij}_{L'L}$ by $\bar{T}^\alpha_{lm+\alpha}$. If
$A'=M$ do the same
with primed indices.}

{\bf 2.} {\em If $A=E$ then replace $L$ in $g^{ij}_{L'L}$ by
$L=l+p,m+\alpha$ and
multiply $g^{ij}_{L'L}$ by $\bar{C}^\alpha(l,p,m)$.
If $A'=E$ do the same with primed indices.}

{\bf 3.} {\em Take the sum over $\alpha=-1,\,0,\,1$,
and (if any) over $p,\,p'=-1,\,1$.}

\vspace*{0.3cm}

The {\em basic
photonic MST equations} (\ref{big1}) are written then as follows:
\begin{equation}
C^o_{iAL} =\sum_{jA'L'}\left[\delta_{ij}\delta_{LL'}\delta_{AA'}
 - \sum_{A"L"} G^{ij}_{AL,A"L"}t^j_{A"L",A'L'}\right]C_{jA'L'}.
\label{mst}
\end{equation}
According to the construction of $G^{ij}$ the sum over $j$
runs here over all $j\neq i$.
Note that $G^{ij}_{AL,A'L'}$ {\em is not} Hermitian. This is a
consequence of using the $t$ matrix which also is not
Hermitian \cite{PW,KKR}. For spherically symmetric scatterers
the $t$ matrix is {\em diagonal} and for
homogeneous spheres it is explicitly known as the {\em Mie solution}
\cite{RN}.

Provided scatterers are identical and arranged in a {\em periodic} way one
can take the Fourier transform with regard to the Bloch momentum ${\bf k}$.
The condition of the existence of a solution to
(\ref{mst}) for $C^o_{iAL}=0$,
\begin{equation}
\det\, \left|\delta_{LL'}\delta_{AA'} -
\sum_{A"L"} G_{AL,A"L"}({\bf k})\,
t_{A"L",A'L'}\right| =0,
\label{kkre}
\end{equation}
then gives the {\em photonic KKR equation}.
It preserves its distinguished feature
known in the case of electrons that is the separation of {\em pure
geometrical} and {\em scattering} properties of a medium.
Geometrical properties are encoded in
{\em geometrical structure constants} characteristic for the
lattice under consideration. They are functions of
energy $\sigma$ and the Bloch momentum ${\bf k}$.
Scattering properties are as usually encoded in {\em phase shifts}
of a {\em single} scatterer.

Our result for the structure constants (\ref{structure})
 essentially {\em disagrees} with Ref. \cite{XZ} : our expression is much
more symmetric while in  Ref. \cite{XZ}
there is no summation over $p,\,p'$ which is necessary here
[see (\ref{vectorgreen})]
since electric multipoles have nonzero matrix elements
with $Y_{L'}$ only for $l'=l\pm 1$
as it is seen from Eq. (\ref{causecl}) of  Appendix \ref{basic}.
This is probably the result that, in contrast to them, we have
kept  trace whether inward or outward values of fields are taken
on the surface of spheres which is necessary to do
for electromagnetic waves \cite{AM}.
Moreover, there are no $\bar{T}_L$ factors in the Eq. (27) of Ref. \cite{XZ}
while their presence follows directly from (\ref{causetl}) of
Appendix \ref{basic}. To visualize the differences one defines
matrices with the entries being column vectors,
namely, the diagonal matrix ${\protect\bf \tau}$,
\begin{equation}
{\protect\bf \tau}_{LL} = [\bar{T}^{-1}_L, \bar{T}^0_L, \bar{T}^1_L],
\end{equation}
and the off-diagonal matrix ${\bf c}$,
\begin{equation}
{\bf c}_{lm;l-1,m}= [\bar{C}^{-1}(l,-1,m), \bar{C}^0(l,-1,m),
\bar{C}^1(l,-1,m)],
\end{equation}
\begin{equation}
{\bf c}_{lm;l+1,m}= [\bar{C}^{-1}(l,1,m), \bar{C}^0(l,1,m),
\bar{C}^1(l,1,m)],
\end{equation}
all other entries being zero.
According to relations (\ref{rel1}) and (\ref{rel2}) of Appendix \ref{basic}
these matrices  are like {\em real unitary} (orthonormal) matrices.
In particular,
\begin{equation}
{\protect\bf \tau}\cdot{\bf \tau}^\dagger
={\bf c}\cdot{\bf c}^\dagger={\bf 1},
\end{equation}
where ${\bf \tau}^\dagger$ and ${\bf c}^\dagger$ are Hermitian
conjugate matrices.
The word ``{\em like}" above is {\em necessary} because the entries
of these matrices are column vectors
and not all the rules of matrix algebra, in particular the cyclicity of
trace, are valid for them. The transformation from scalar to
vector structure constants can be then written in a matrix form as
\begin{equation}
G^{ij}=
\left(
\begin{array}{cc}
\tau &0\\
0& {\bf c}
\end{array}
\right)
 \left(
\begin{array}{cc}
\tilde{g}^{ij} & \tilde{g}^{ij}\\
 \tilde{g}^{ij}&  \tilde{g}^{ij}
\end{array}
\right)^{\mbox{tr}}
 \left(
\begin{array}{cc}
{\protect\bf\tau }&0\\
0& {\bf c}
\end{array}
\right)^\dagger
,
\label{mform}
\end{equation}
where ``tr" means the transposition of $AL$ and $A'L'$ indices. Here
the matrix $\tilde{g}^{ij}$ has the entries diagonal $3\times 3$ matrices,
\begin{equation}
\tilde{g}^{ij}_{LL'}= \left(
\begin{array}{ccc} g^{ij}_{lm-1;l'm'-1} &0&0\\
0& g^{ij}_{lm;l'm'}&0\\
0&0&g^{ij}_{lm+1;l'm'+1}
\end{array}
\right).
\label{gform}
\end{equation}
Provided $|m\pm 1|>l$ or $|m'\pm 1|>l'$ the entry in this matrix
can be set in principle to be any finite
number since in that case it is multiplied by the zero element of
either the ${\bf \tau}$ or ${\bf c}$ matrix.

\section{Operator formalism for the Maxwell equations}
\label{oper}
In the case of the Schr\"{o}dinger operator $\mbox{H}_o$
with some potential
$\Gamma$ it is common to suppress the spatial indices and to consider
all the quantities including the Green functions as operators
in the Hilbert space. For example the DOS is then given
by the formula
\begin{equation}
\rho(E)=-\frac{1}{\pi}\mbox{ImTr}\,G(E_+)
=-\frac{1}{\pi}\mbox{ImTr}\,\frac{1}{E_+-\mbox{H}},
\label{dos}
\end{equation}
where $\mbox{H}=\mbox{H}_o+\Gamma$ and $E_+$ stands for the limit
$s\rightarrow 0_+$ in $E=E+is$.

In the case of the Maxwell equation problem is a bit more subtle.
The Maxwell equation (\ref{ME*}) can be written formally as
\begin{equation}
\tilde{\mbox{H}}\,|{\bf E}\rangle
=-(\mbox{\bf 1}\,\mbox{\boldmath $\nabla$}^2 -
\mbox{\boldmath $\nabla$}\otimes\mbox{\boldmath $\nabla$})|{\bf E}\rangle
=
\omega^2\epsilon\,|{\bf E}\rangle.
\label{MEO}
\end{equation}
Note that as the consequence of
$\mbox{\boldmath $\nabla$}\tilde{\mbox{H}}\equiv 0$
any solution of (\ref{MEO}) automatically satisfies
$\mbox{\boldmath $\nabla$}\cdot(\epsilon\,|{\bf E}\rangle)
=\mbox{\boldmath $\nabla$}\cdot|{\bf D}\rangle=0$,
the ``transversality" condition in the case of spatially varying
$\epsilon({\bf r})$.
Equation \ (\ref{MEO}) resembles an eigenvalue equation, however, the
`eigenvalue' $\omega^2$ is multiplied by the spatially varying function
$\epsilon({\bf r})$. If one defines the Green function as
\begin{equation}
\tilde{G}(\omega^2)=\frac{1}{\omega^2\epsilon\,
\mbox{\bf 1}+\mbox{\bf 1}\,\mbox{\boldmath $\nabla$}^2 -
\mbox{\boldmath $\nabla$}\otimes\mbox{\boldmath $\nabla$}}
\label{gbar}
\end{equation}
one would find that $\tilde{G}$ is {\em not diagonal}
in the basis of eigenstates $|n\rangle$ of the frequency $\omega_n$
and $\mbox{ImTr}\,\tilde{G}(\omega^2+is)$ {\em is not} proportional to the
level density $\rho(\omega^2)$,
\begin{equation}
-\frac{1}{\pi}\lim_{s\rightarrow 0_+}
\mbox{ImTr}\,\tilde{G}(\omega^2+is)=\sum_n\langle n|\,\epsilon^{-1}\,|n\rangle
\delta(\omega^2-\omega^2_n).
\end{equation}
The {\em right} resolvent which gives correctly the DOS via (\ref{dos})
and is diagonal in the ``energy representation" is
\begin{equation}
G(\omega^2)=\frac{1}{\omega^2\,\mbox{\bf 1}+
\epsilon^{-1}(\mbox{\bf 1}\,\mbox{\boldmath $\nabla$}^2 -
\mbox{\boldmath $\nabla$}\otimes\mbox{\boldmath $\nabla$})}=
\tilde{G}(\omega^2)\,\epsilon.
\end{equation}
For our purposes it is convenient to define the parameter $E$ in the case
of photons as $E=\sigma^2=\omega^2\epsilon_o={\bf k}^2$, where ${\bf k}$
is the wave vector in a medium with permittivity $\epsilon_o$. As
usually, $E={\bf p}^2/2m$ for electrons.
The above considerations then imply that the {\em right} analog
of $\mbox{H}$ (Ref. \cite{Ham}) in the measure $dE$ is
\begin{equation}
\mbox{H}=-\frac{\epsilon_o}{\epsilon}(\mbox{\bf 1}\,
\mbox{\boldmath $\nabla$}^2
-\mbox{\boldmath $\nabla$}\otimes\mbox{\boldmath $\nabla$}).
\label{hami}
\end{equation}
The {\em right} separation of $\mbox{H}$ as $\mbox{H}=\mbox{H}_o+\Gamma$
for the Maxwell equation then reads
\begin{eqnarray}
\mbox{H}_o&=&-(\mbox{\bf 1}\,\mbox{\boldmath $\nabla$}^2
-\mbox{\boldmath $\nabla$}\otimes\mbox{\boldmath $\nabla$}),
\nonumber\\
\mbox{H}&=&\frac{\epsilon_o}{\epsilon}\,\mbox{H}_o=
\left(1-\frac{\epsilon}{v}\right)\Gamma,
\nonumber\\
\Gamma&=&\mbox{H}-\mbox{H}_o=\left(\frac{\epsilon_o-\epsilon}{\epsilon}
\right)\mbox{H}_o=
-\frac{v}{\epsilon}\,\mbox{H}_o=-\frac{v}{\epsilon_o}\,\mbox{H}.
\label{decom}
\end{eqnarray}
Despite the formal similarity of (\ref{ME*}) with the Schr\"{o}dinger
equation note the principal difference: the differential operators are
generically multiplied by spatially varying functions
[see (\ref{hami}) and (\ref{decom})]. All that is the consequence of
(\ref{ME*}) where the `eigenvalue' $\omega^2$ multiplies the potential
$v({\bf r})$, and, therefore, this feature persists for any differential
equation with `energy dependent' potential.
In the language of the Schr\"{o}dinger equation the problem
of finding the spectrum of (\ref{ME*}) reads as follows : look
for {\em different} Hamiltonians $\mbox{H}_\omega$ with the potential
$\omega^2\epsilon({\bf r})$ and find all $\omega$ such that
$\mbox{H}_\omega$ has zero in its spectrum.
The price we pay for {\em removing} the energy dependence
is the introduction of potential $\Gamma$ which is itself a
differential operator.
However, because of (\ref{decom}), the potential $\Gamma$
reduces to a {\em purely multiplicative} operator on eigenfunctions of
both $\mbox{H}_o$ and $\mbox{H}$.

Let us now find the free Green function $G_o(z)=(z-\mbox{H}_o)^{-1}$
for electromagnetic waves. We shall show that
\begin{equation}
G_o(z)=
\left[\mbox{\bf 1}+\frac{1}{z}
\mbox{\boldmath $\nabla$}\otimes\mbox{\boldmath $\nabla$}\right]
\frac{1}{z+\mbox{\boldmath $\nabla$}^2}
=-\frac{1}{\omega^2}{\bf d}_o
\hspace*{1cm}\mbox{and}\hspace*{1cm}
\mbox{\boldmath $\nabla$}
\cdot G_o(z)=\frac{1}{z}\mbox{\boldmath $\nabla$}
\cdot
\label{gof}
\end{equation}
The Green function
${\bf d}_o$ here is the same as defined in (\ref{d0}),
\begin{equation}
{\bf d}_o=-\left[\omega^2\mbox{\bf 1}+\frac{1}{\epsilon_o}
\mbox{\boldmath $\nabla$}\otimes\mbox{\boldmath $\nabla$}\right]
\frac{1}{\sigma^2+\mbox{\boldmath $\nabla$}^2}\cdot
\end{equation}
Obviously, Eq.\ (\ref{gof}) is equivalent to
\begin{equation}
-\mbox{H}_o {\bf d}_o=
(\mbox{\bf 1}\,\mbox{\boldmath $\nabla$}^2 -
\mbox{\boldmath $\nabla$}\otimes\mbox{\boldmath $\nabla$}) {\bf d}_o
 = -\omega^2\,\mbox{\bf 1}-\omega^2\epsilon_o{\bf d}_o
\hspace*{1cm}\mbox{and}\hspace*{1cm}
\mbox{\boldmath $\nabla$}\cdot{\bf d}_o=-\frac{1}{\epsilon_o}
\mbox{\boldmath $\nabla$}\cdot
\label{rel}
\end{equation}
To prove it one uses
\begin{equation}
\mbox{\boldmath $\nabla$}^2\frac{1}{\sigma^2 +
\mbox{\boldmath $\nabla$}^2} =
1-\sigma^2\frac{1}{\sigma^2+\mbox{\boldmath $\nabla$}^2}
\end{equation}
to show that the contributions of
\begin{equation}
\mbox{\boldmath $\nabla$}^2
\left(\mbox{\boldmath $\nabla$}\otimes\mbox{\boldmath $\nabla$}
\frac{1}{\sigma^2+\mbox{\boldmath $\nabla$}^2}\right) =
\mbox{\boldmath $\nabla$}\otimes\mbox{\boldmath $\nabla$} -
\sigma^2\mbox{\boldmath $\nabla$}\otimes\mbox{\boldmath $\nabla$}
\frac{1}{\sigma^2+\mbox{\boldmath $\nabla$}^2}
\end{equation}
and
\begin{equation}
-\mbox{\boldmath $\nabla$}\otimes\mbox{\boldmath $\nabla$}
\left(\mbox{\boldmath $\nabla$}\otimes\mbox{\boldmath $\nabla$}
\frac{1}{\sigma^2+\mbox{\boldmath $\nabla$}^2}\right) =
-\mbox{\boldmath $\nabla$}\otimes\mbox{\boldmath $\nabla$} +
\sigma^2\mbox{\boldmath $\nabla$}\otimes\mbox{\boldmath $\nabla$}
\frac{1}{\sigma^2+\mbox{\boldmath $\nabla$}^2}
\end{equation}
cancel. Since
\begin{equation}
(\mbox{\bf 1} \mbox{\boldmath $\nabla$}^2 -
\mbox{\boldmath $\nabla$}\otimes\mbox{\boldmath $\nabla$})
\left(\omega^2\frac{1}{\sigma^2+\mbox{\boldmath $\nabla$}^2}\right)
= \omega^2\mbox{\bf 1}+\omega^2\epsilon_o{\bf d}_o,
\end{equation}
one easily shows the first of relations (\ref{rel}).
As for the second one, note that
\begin{equation}
\mbox{\boldmath $\nabla$}\cdot{\bf d}_o
=-\mbox{\boldmath $\nabla$}\left[\omega^2\mbox{\bf 1}+
\frac{1}{\epsilon_o}\mbox{\boldmath $\nabla$}^2\right]
\frac{1}{\sigma^2 +\mbox{\boldmath $\nabla$}^2}
=-\frac{1}{\epsilon_o}\mbox{\boldmath $\nabla$}\cdot
\end{equation}

It is worthwhile to see  how the Lippmann-Schwinger equation
(\ref{entq1}) of the preceding section
works within the operator formalism.
Equation (\ref{entq1}) is written now as
\begin{equation}
|{\bf E}\rangle = |{\bf E}_o\rangle -G_o\omega_n^2 v|{\bf E}\rangle,
\label{operls}
\end{equation}
where $v=\epsilon-\epsilon_o$,
$|{\bf E}_o\rangle$ is the eigenfunction of $\mbox{H}_o$
corresponding to an eigenenergy $E_n=\omega^2_n\epsilon_o$,
 and $G_o=G_o(E_n+i0)$.
This equation is seemingly in contradiction with
the {\em standard} form of the Lippmann-Schwinger
equation  \cite{PW} for electric intensity $|{\bf E}\rangle$
with regard to the decomposition (\ref{decom})
of $\mbox{H}$ as $\mbox{H}=\mbox{H}_o+\Gamma$
which implies it to be
\begin{equation}
|{\bf E}\rangle = |{\bf E}_o\rangle + G_o\Gamma\, |{\bf E}\rangle.
\label{operls1}
\end{equation}
Formally, since $\mbox{H}_o G_o(z)=-1+zG_o(z)$,
\begin{equation}
\mbox{H}_o |{\bf E}\rangle
= E_n |{\bf E}_o\rangle - \Gamma\,|{\bf E}\rangle
+E_n G_o\Gamma\, |{\bf E}\rangle.
\end{equation}
By iterating this equation with the help of Eq.\ (\ref{operls1})
one generates the Rayleigh-Schr\"{o}dinger perturbation
series \cite{Kat}. After their summation one obtains then the
desired result,
\begin{eqnarray}
\mbox{H}\,|{\bf E}\rangle =(\mbox{H}_o  +\Gamma)|{\bf E}\rangle =
E_n(1+ G_o\Gamma+
G_o\Gamma G_o\Gamma +\ldots)|{\bf E}_o\rangle
\nonumber\\
=E_n\frac{1}{1- G_o\Gamma}\,  |{\bf E}_o\rangle=
E_n |{\bf E}\rangle.
\label{inter*}
\end{eqnarray}
To show that (\ref{operls})
 works as well note that
\begin{equation}
\mbox{H}_o |{\bf E}\rangle
= E_n |{\bf E}_o\rangle + \omega_n^2 v|{\bf E}\rangle
-E_n G_o\omega_n^2 v|{\bf E}\rangle.
\end{equation}
Similarly as above one iterates this equation, now with the help of
(\ref{operls}), and  finds
\begin{eqnarray}
(\mbox{H}_o  -\omega_n^2v)|{\bf E}\rangle =
E_n(1- G_o\omega_n^2 v+
G_o\omega_n^2 vG_o\omega_n^2 v -\ldots)|{\bf E}_o\rangle
\nonumber\\
=E_n\frac{1}{1+G_o\omega_n^2v}\,  |{\bf E}_o\rangle=
E_n |{\bf E}\rangle
\label{inter}
\end{eqnarray}
that is equivalent to (\ref{MEO}).

To show the `transversality' of the solutions note that by applying
$\mbox{\boldmath $\nabla$}\cdot$ on (\ref{operls1}) one finds by using
the second of relations (\ref{gof})
\begin{equation}
\mbox{\boldmath $\nabla$}\cdot|{\bf E}\rangle =
\mbox{\boldmath $\nabla$}\cdot|{\bf E}_o\rangle
+ \frac{1}{z}
\mbox{\boldmath $\nabla$}\cdot(\Gamma\,|{\bf E}\rangle).
\end{equation}
Now
\begin{equation}
\Gamma\, |{\bf E}\rangle=-\frac{v}{\epsilon_o}
\mbox{H}\, |{\bf E}\rangle=-\omega_n^2v |{\bf E}\rangle,
\end{equation}
and
\begin{equation}
\mbox{\boldmath $\nabla$}\cdot(\epsilon\,|{\bf E}\rangle) =
\epsilon_o \mbox{\boldmath $\nabla$}\cdot|{\bf E}_o\rangle.
\end{equation}
Similarly, one shows the `transversality' for the solutions of
(\ref{operls}),
\begin{equation}
\mbox{\boldmath $\nabla$}\cdot|{\bf E}\rangle =
\mbox{\boldmath $\nabla$}\cdot|{\bf E}_o\rangle-\frac{1}{\epsilon_o}
\mbox{\boldmath $\nabla$}\cdot(v|{\bf E}\rangle),
\hspace*{0.5cm}\mbox{i.e.,}\hspace*{0.5cm}
\mbox{\boldmath $\nabla$}\cdot(\epsilon\,|{\bf E}\rangle) =
\epsilon_o \mbox{\boldmath $\nabla$}\cdot|{\bf E}_o\rangle.
\end{equation}
Thus if one takes $|{\bf E}_o\rangle$ to be `transverse' then
$|{\bf E}\rangle$ will be `transverse' too.
In other words, both, operator $(1- G_o\Gamma)^{-1}$ and
 $(1+G_o\omega^2 v)^{-1}$, map `{\em transverse}'
eigenfunctions of $\mbox{H}_o$ onto `{\em transverse}' eigenfunctions of
$\mbox{H}$.
Provided the operator $(1-G_o\Gamma)^{-1}$  or
 $(1+G_o\omega^2 v)^{-1}$  exists
then $\mbox{H}$ has an  eigenfunction $|{\bf E}\rangle$,
\begin{equation}
|{\bf E}\rangle = \frac{1}{1-G_o\Gamma}\, |{\bf E}_o\rangle
= \frac{1}{1+G_o\omega^2 v}\, |{\bf E}_o\rangle ,
\end{equation}
of the same eigenenergy $E_n$ as $|{\bf E}_o\rangle$,
which can be constructed by the Fredholm method \cite{RN}.
One can show that $[1+G_o(E_n+i0)\omega^2_n v]^{-1}$ coincides
with the restriction
of $[1- G_o(E_n+i0)\Gamma]^{-1}$ on the space of the eigenfunctions
of $\mbox{H}_o$ with
eigenenergy $E_n=\omega_n^2\epsilon_o$. By using the identity
$[1- G_o\Gamma]^{-1}=1+G\Gamma$  one finds
\begin{equation}
\frac{1}{1-G_o(E_n+i0)\Gamma} |{\bf E}_o\rangle= [1+ G(E_n+i0)
\Gamma]  |{\bf E}_o\rangle=
[1-\tilde{G}(\omega^2_n+i0)\omega_n^2v] |{\bf E}_o\rangle,
\end{equation}
where $\tilde{G}(\omega^2_n+i0)$ is defined by (\ref{gbar}).
On the other hand,
\begin{equation}
\frac{1}{1+G_o\omega^2_n v}
= 1-\frac{1}{\omega_n^2\epsilon_o+i0-(\mbox{H}_o-\omega^2_nv)}
\omega_n^2 v =1-\tilde{G}(\omega_n^2+i0) \omega_n^2 v.
\end{equation}

So far we have not addressed the question of convergence of
the Rayleigh-Schr\"{o}dinger perturbation series that
occur, for example, in (\ref{inter*}) and (\ref{inter}). For those who
are interested in this question we refer to Ref. \cite{Mor} where
an improvement of the Rayleigh-Schr\"{o}dinger perturbation theory
is given by the help of a generalization of the Borel summability
method.

After the decomposition (\ref{decom}) one can repeat all standard
techniques known for the Schr\"{o}dinger equation.
One defines the $T$ matrix
by the Lippmann-Schwinger equation,
\begin{equation}
T:=\Gamma+\Gamma G_o^+ T
=\Gamma+TG_o^+\Gamma =\Gamma (1-G_o^+\Gamma)^{-1}
=(1-\Gamma G_o^+)^{-1}\Gamma,
\label{tmatrix}
\end{equation}
with  $G_o^+$ satisfying the outgoing boundary conditions.
Sometimes it is more convenient to work with
{\em Hermitian} quantity as the $K$ matrix. The $K$ matrix is
defined essentially in the same manner as the $T$ matrix,
\begin{equation}
K:=\Gamma +\Gamma G_o^oK=\Gamma + KG_o^o\Gamma =
\Gamma(1-G_o^o\Gamma)^{-1}=(1-\Gamma G_o^o)^{-1}\Gamma.
\label{kmatrix}
\end{equation}
The Green function $G_o^o$ here is, however, the
{\em real} or the {\em Hermitian} part of $G_o^+$,
\begin{equation}
G_o^+=G_o^o-iD,
\label{deco}
\end{equation}
and $D$ is a solution of the homogeneous equation.
The $T$ matrix can be expressed in terms of $K$ matrix and vice versa,
\begin{equation}
T=K(1+iD K)^{-1},\hspace*{1cm}K=T(1-iD T)^{-1}.
\label{ktrelation}
\end{equation}
For a spherically symmetric scatterer
the channel  $\mbox{\bf S}_L$ matrix can be expressed as
\begin{equation}
\mbox{\bf S}_L=\frac{1-iK_L}{1+i K_L}=1-2iT_L=e^{2i\eta_l},
\label{smatrix}
\end{equation}
where
\begin{equation}
T_L=-\sin\eta_l\, e^{i\eta_l}.
\label{tsmatrix}
\end{equation}
Here $\eta_l$ is the phase shift, which in this special case, does not
depend on the magnetic quantum number \cite{PW}.

\section{Density of states calculations}
\label{path}
We now turn on to the calculation of DOS's in a system of
nonoverlapping scattering centers. In the next we shall
follow the original calculations of Lloyd and Smith
for the DOS of electrons \cite{PW}. When considering
the thermodynamic limit $\Omega\rightarrow\infty$ and $N\rightarrow\infty$
such that the density of scatterers $N/\Omega$ stays finite one can
take one of two fundamentally different points of view:
either the system has an infinity volume $\Omega$ and thus occupies
all space, or,
the system is situated in a much larger volume $\Omega_\infty$
with the limits being taken so that
$\Omega_\infty\rightarrow\infty$ before $\Omega\rightarrow\infty$ and
$\Omega/\Omega_\infty\rightarrow 0$
even though the volume $\Omega$ becomes infinite. In the first case
the integrated density of states (IDOS's) $N(E)$ is found from the formula
\begin{equation}
N(E)=-\frac{1}{\pi}\mbox{ImTr}\ln[E_+ -\mbox{H}]=
N_o(E)-\frac{1}{\pi}\mbox{ImTr}\ln[1-G_o^+\Gamma],
\label{doper}
\end{equation}
with $\Gamma=\sum_i \Gamma_i$, where the sum runs over
all scatterers \cite{PW}.
In the second case the IDOS is determined directly
by the {\bf S} matrix of the system via the Krein-Friedel formula
\begin{equation}
N(E)=N_o(E)+\frac{1}{2\pi i}\mbox{Tr}\ln\mbox{\bf S}=N_o(E)+
\frac{1}{2\pi}\mbox{ImTr}\ln\mbox{\bf S}.
\label{ds}
\end{equation}

To calculate the change of the IDOS induced by the presence of scatterers
and establish the equivalence of (\ref{doper}) and
(\ref{ds}) one expands the logarithm in (\ref{doper}),
\begin{equation}
\triangle N(E)=-\frac{1}{\pi}\mbox{ImTr}\ln\left[1-G_o^+\Gamma\right]=
\frac{1}{\pi}\mbox{ImTr}\sum_{n=1}^\infty \frac{1}{n}(G_o^+\Gamma)^n,
\label{dexp}
\end{equation}
and substitutes $\Gamma=\sum_i \Gamma_i$.
A generic term of the expansion is given by
\begin{equation}
\frac{1}{n}\mbox{Tr}\, (A_{i_1}A_{i_2}\ldots A_{i_n})
\label{term}
\end{equation}
with $A_{i_k}=G_o^+\Gamma_{i_k}$ where the indices $i_k$
may equal. The diagram corresponding to this term is constructed
by connecting vertices $i_1$, $i_2$, $\ldots$, $i_n$ in
subsequent order.
Because of the trace operation the vertices $i_n$ and $i_1$
are connected, too. If some pair of vertices is equal one
draws a {\em tadpole} which starts and terminates at
the vertex (see Fig.\ \ref{figtad}).
\begin{figure}
\centerline{\epsfxsize=8cm \epsfbox{flfig1.ps}}
\caption{Single-site tadpole diagrams.}
\label{figtad}
\end{figure}
If in some $p$-tuple of vertices all the vertices
have the same label then $p-1$ tadpoles are drawn from
the given vertex. To each vertex $i_k$ corresponds the scattering
potential $\Gamma_{i_k}$ and to each line connecting subsequent
vertices the propagator $G_o^+(i_k-i_{k+1})$.

In what follows it is convenient to imagine the term (\ref{term}) as a
{\em closed trajectory}. It is clear that
any closed trajectory of $n$th order with at least
two different vertices occurs exactly $n$ times
in the expansion (\ref{dexp}):
any term  $A_{i_1}A_{i_2}\ldots A_{i_n}$ with cyclic permutation of
indices $i_1 i_2\ldots i_n$  gives the same trajectory.
The main reason to speak about the trajectories is that sometimes
{\em different} closed trajectories leading to the {\em same}
diagram may give a {\em different} contribution.
An example of this are
trajectories $(1,2,3,2,3,4,3,2)$ and $(1,2,3,4,3,2,3,2)$
(see Fig.\ \ref{fig2}).
They  are
different modulo cyclic permutation of indices, nonetheless
they give rise to the same diagram.
\begin{figure}
\centerline{\epsfxsize=8cm \epsfbox{flfig2.ps}}
\caption{Two different closed trajectories which give different
contributions to $\triangle N(E)$ but lead to the same diagram. The
numbers indicate successive paths between different scatterers from
which given closed  trajectories are composed.}
\label{fig2}
\end{figure}
If one then follows
the rules to calculate the contribution of the trajectories
one finds that they are different if the scatterers are not
identical or not of the same distance each from other
since then
\begin{equation}
\mbox{Tr}\,(A_1A_2A_3A_2A_3A_4A_3A_2)\neq
\mbox{Tr}\,(A_1A_2A_3A_4A_3A_2A_3A_2),
\end{equation}
the fact that seems to have been unnoticed so far.

As shown by Lloyd and Smith \cite{PW} all {\em tadpole} diagrams
that give multiple scattering from the same
site  can be summed over. One first resums
the tadpole diagrams with
a single site (see Fig.\ \ref{figtad}).
The result for the site $j$ is simply
\begin{equation}
-\frac{1}{\pi}\mbox{ImTr}\ln\left[1-G_o^+\Gamma_{j}\right].
\end{equation}
It can be rewritten to the more familiar Friedel form
\begin{equation}
-\frac{1}{\pi}\mbox{ImTr}\ln\left[1-G_o^+\Gamma_{j}\right]=
\frac{1}{2\pi}\mbox{ImTr}\ln\mbox{\bf S}^{j}=\frac{1}{\pi}\sum_{AL}
\eta^{j}_{AL,AL},
\label{friedel}
\end{equation}
where {\bf S}${}^{j}$ is the single site {\bf S} matrix on the site $j$,
and $\eta^j_{AL,A'L'}$ are the corresponding phase shifts \cite{PW}.
One obtains the result under the hypothesis that
$(1-G_o^o \Gamma_{j})$ has no zeros and poles on the
real axis.
It is known that if $F(E)$ is an analytical function in a strip
$E+is$ of the upper half-plane, which is real when $E$ is real and which
has zeros $E_z$ and poles $E_p$ on the real axis, then
\begin{equation}
\lim_{s\rightarrow 0}\mbox{Im}\ln F(E+is)= -\pi\sum_z \Theta(E-E_z)+
\pi\sum_p \Theta(E-E_p),
\label{detfo}
\end{equation}
where $\Theta(x)$ is the Heaviside step function \cite{MB}.
Thus, if the above hypothesis is satisfied then
$\mbox{ImTr}\ln\left[1-G_o^o \Gamma_{j}\right]=0$ and
\begin{equation}
-\frac{1}{\pi}\mbox{ImTr}\ln\left[1-G_o^+\Gamma_{j}\right]=
-\frac{1}{\pi}\mbox{ImTr}\ln\left[1+iD k^j\right],
\label{kform}
\end{equation}
where $k^j$ is the single-site $k$ matrix.
In the angular momentum representation  \cite{PW,AMo}
\begin{equation}
\left.D_{AL,A'L'}({\bf R})\right|_{{\bf R}=0} =\delta_{LL'}\delta_{AA'}.
\label{drel}
\end{equation}
Since $k_{AL,A'L'}^j$ is real, $\mbox{Im}\ln[1-i k_{AL,A'L'}^j]=-
\mbox{Im}\ln[1+i k_{AL,A'L'}^j]$,
and using the relation (\ref{smatrix}) one finally establishes
(\ref{friedel}). Therefore, the change of the IDOS can be expressed as
\begin{equation}
\triangle N(E)= \frac{1}{\pi} \sum_{j=1}^N\sum_{AL}\eta^j_{AL,AL} +
\triangle N^{(1)}(E),
\label{sep}
\end{equation}
where $\triangle N^{(1)}(E)$ is entirely due to the presence of
{\em multiple scatterers} and {\em multiple-scattering effects}.
In order to calculate $\triangle N^{(1)}(E)$ one sums
over tadpoles in the remaining diagrams.
The summation results in replacing the vertex contribution
$\Gamma_j$ by the single-site $t$ matrix $t^j$,
\begin{equation}
\Gamma_j\rightarrow t^j=\Gamma_j\frac{1}{1-G_o^+\Gamma_j}\cdot
\end{equation}
For example, for two different vertices $i_1=1$ and
$i_2=2$ one has
\begin{equation}
\mbox{Tr}\sum_{m=1}^\infty\sum_{n=1}^\infty
G_o^+\Gamma_1(G_o^+\Gamma_1)^{m-1}
G_o^+\Gamma_2
(G_o^+\Gamma_2)^{n-1}=\mbox{Tr}\,G_o^+\Gamma_1
\frac{1}{1-G_o^+\Gamma_1}G_o^+\Gamma_2
\frac{1}{1-G_o^+\Gamma_2}\cdot
\end{equation}
$\triangle N^{(1)}(E)$ is then written as
$$
\triangle N^{(1)}(E) =  \frac{1}{\pi}\mbox{ImTr} \left(
 \frac{1}{2} \sum_i\sum_{j\neq i} G_o^+ t^i G_o^+ t^j +\frac{1}{3}
\sum_i \sum_{j\neq i}\sum_{k\neq i,j}G_o^+ t^iG_o^+ t^j G_o^+ t^k
+ \ldots\right)
$$
where any two subsequent (with regard to the trace) $t$ matrices
have a different label.

By virtue of the nonoverlapping condition the trace
can be taken either in the basis of spherical harmonics
in the case of electrons, or, of electric and magnetic multipoles
in the case of photons.
Because of the trace operation any closed trajectory
becomes a closed trajectory in the symbolic {\em `phase space'} and
the {\em diagrammatic rules} to calculate $\triangle N^{(1)}(E)$
can be formulated as follows on the energy shell :

\vspace*{0.3cm}

{\em
{\bf 1.} Draw all possible closed  trajectories-orbits.
Any intermediate path has to connect different sites but it can return
back to the original site after visiting some different site.

{\bf 2.}
To any orbit corresponds a diagram with vertices given
by the single-site $t$ matrices $t_{AL,A'L'}^j$.
To the line
connecting the  $j$th  and $n$th sites corresponds
the propagator matrix $G^{jn}_{AL,A'L'}=
G_{AL,A'L'}^+({\bf R}_n -{\bf R}_j)$.

{\bf 3.}
The contribution of a given orbit  is obtained
by taking the trace of a matrix which results by multiplying
$t$ and $G$ matrices in the order that is determined by tracing the
orbit.

{\bf 4.}
The total contribution to the density of states is given by
summing over all orbits and by adding the Friedel sum
(\ref{friedel}) for each scatterer.}

\vspace*{0.3cm}

All the above rules can be summarized by the analytical expression
(cf. Ref. \cite{PW})
\begin{eqnarray}
\lefteqn{\triangle N(E)= \frac{1}{\pi}\sum_{j=1}^N\sum_{AL}\eta^j_{AL,AL}-
}\nonumber\\
&&-\frac{1}{\pi}\mbox{ImTr}\ln
\left[\delta_{LL'}\delta_{AA'}\delta_{jn}-\sum_{A_1L_1}{\cal
G}_{AL,A_1L_1}^{jn}t_{A_1L_1,A'L'}^n\right],
\label{dfinal}
\end{eqnarray}
where $\sum_{A_1L_1}{\cal G}_{AL,A_1L_1}^{jn}t_{A_1L_1,A'L'}^n$ is the
matrix with three pairs of indices :
angular momentum $LL'$, multipole $AA'$, and spatial ones
$jn$. The trace is then taken over all the pairs of indices.
The Green function (propagator) ${\cal G}_{AL,A'L'}^{jn}$ here equals
$G_{AL,A'L'}^{+jn}$ except for $j=n$ where it is zero.
It is the familiar structure constant of Sec. \ref{mstsec}.
Formally, one can write
\begin{equation}
{\cal G}_{AL,A'L'}^{jn}=\alpha_j G_{AL,A'L'}^{+jn}
\label{form}
\end{equation}
where $\alpha_j$ is a `Grassmann-like' variable on the site $j$
(Grassmann-like because $\alpha_j^2=0$ but neither commutation
nor anticommutation of $\alpha_j$ and $\alpha_n$ are defined).

Sometimes it is more convenient
to rewrite (\ref{dfinal}) to a slightly different form
in terms of $k$ matrix.
By combining (\ref{friedel}) and (\ref{kform}) one
can write
\begin{equation}
\frac{1}{\pi}\sum_{AL} \eta^j_{AL,AL}=
-\frac{1}{\pi}\mbox{ImTr}\ln\left\{\left[\delta_{LL'}\delta_{AA'}
+i k_{AL,A'L'}^j\right]\delta_{jn}\right\}.
\end{equation}
Afterwards both term in (\ref{dfinal}) are written in the similar
matrix form, one can sum the logarithms by multiplying
their arguments as matrices, and obtain
\begin{eqnarray}
\lefteqn{
-\frac{1}{\pi}\mbox{ImTr}\ln\left\{
\delta_{LL'}\delta_{AA'}\delta_{jn}+ik_{AL,A'L'}^j-
\sum_{ALA_2L_2}{\cal G}_{AL,A_2L_2}^{jn}t_{A_2L_2,A'L'}^n\right.}
\hspace{3cm}\nonumber\\
&&
\left.-
i\sum_{A_1L_1A_2L_2}k_{AL,A_1L_1}^j{\cal G}_{A_1L_1,A_2L_2}^{jn}
t_{A_2L_2,A'L'}^n\right\}.
\end{eqnarray}
After using the relations (\ref{ktrelation}), (\ref{drel}), (\ref{form}) and
the cyclicity property of trace one arrives at
\begin{equation}
\triangle N(E)=-\frac{1}{\pi}\mbox{ImTr}\ln
\left[\delta_{LL'}\delta_{AA'}\delta_{jn}-
\sum_{A_1L_1}G_{AL,A_1L_1}^{+jn}k_{A_1L_1,A'L'}^n\right].
\label{dreal}
\end{equation}

The formalism is well suited for a {\em finite cluster}
of scatterers. If the number $N$ of scatterers tends to
infinity one defines the integrated density of states
$\bar{N}(E)=N(E)/V$ per unit volume.
 After a suitable averaging over the position of scatterers
the formalism has also been applied by Lloyd to describe liquids \cite{LP}.
If the scatterers are {\em identical}
and arranged in a periodic manner a useful tool to perform the summation
over lattice sites in the expansion of (\ref{dfinal}) is to use the
{\em lattice} Fourier transform,
\begin{equation}
{\cal G}_{AL,A_1L_1}({\bf k})=
\sum_{{\bf n}\in\Lambda}
{\cal G}_{AL,A_1L_1}^{jn} e^{i{\bf k}\cdot({\bf n}-{\bf j})}
= \sum_{{\bf n}\in\Lambda}
{\cal G}_{AL,A_1L_1}^{nj} e^{i{\bf k}\cdot({\bf j}-{\bf n})},
\end{equation}
where the sum runs over all points of the lattice $\Lambda$ and
${\bf k}$ is the Bloch momentum. Because of the cyclicity
of the trace operation in (\ref{dfinal}) the trace does not change
under the substitution
\begin{equation}
{\cal G}_{AL,A_1L_1}^{jn} \rightarrow
{\cal G}_{AL,A_1L_1}^{jn} e^{i{\bf k}\cdot({\bf n}-{\bf j})}.
\end{equation}
In summing term by term in the expansion of (\ref{dfinal})
one fixes one lattice index. The summation over
the remaining lattice indices then gives the Fourier transform of
${\cal G}_{AL,A_1L_1}^{jn}$. Eventually, one finds the change of the
integrated density of states $\triangle\bar{N}(E)$
{\em per lattice site} ({\em unit cell volume V }) to be
\begin{equation}
\triangle\bar{N}(E)=\frac{1}{\pi} \sum_{AL}\eta_{AL,AL} (E)
-\frac{1}{\pi}\mbox{ImTr}\ln
\left[\delta_{LL'}\delta_{AA'}-\sum_{A_1L_1}
{\cal G}_{AL,A_1L_1}({\bf k})\,t_{A_1L_1,A'L'}\right]
\label{kkfu}
\end{equation}
[cf. (\ref{kkre})].
Although we have been interested in the three-dimensional MST and KKR
our diagrammatic rules remain intact in two dimensions.
The only change concerns the set of indices over which the trace
is taken. For example, in the case of electrons the multi-index
$L=lm$ is simply reduced to $l$.
As a self-consistency check note that in the case of an {\em empty} lattice
both the phase shifts $\eta_{AL,A'L'}$ and the $t$ matrix
[cf. (\ref{tsmatrix})] are {\em zero}
and hence, according to (\ref{kkfu}), $\triangle\bar{N}(E)=0$.
To get full IDOS $N(E)$ in three dimensions one has to add (\ref{kkfu})
to
\begin{equation}
N_o(E)=\frac{Vp^3}{3\pi^2\hbar^3}=\frac{V(2m E)^{3/2}}{3\pi^2\hbar^3}
\label{form1}
\end{equation}
in the case of electrons, and
\begin{equation}
N_o(E)=\frac{V k^3}{3\pi^2}=\frac{V E^{3/2}}{3\pi^2}
\label{form2}
\end{equation}
in the case of photons.

Apart from the band structure calculations the expression (\ref{kkfu})
for $\triangle\bar{N}(E)$ can be used directly to calculate
the IDOS of a {\em quantum billiard} on a {\em torus} after one has set
${\bf k}=0$ (this is equivalent to impose the periodic boundary conditions).
In the special case of a two-dimensional billiard on a torus the relation
(\ref{kkfu}) gives the result that is equivalent to
Eq. (6.11) of Ref. \cite{MB} without applying (\ref{detfo})
on the KKR determinant (\ref{kkre}) and without
using a particular form of the structure constants.
A useful methodological tool to study the semiclassical DOS
of classically ergodic
systems is the {\em Gutzwiller trace formula} \cite{MG},
\begin{equation}
\mbox{Tr}\,G({\bf r},{\bf r},E)=\sum_j\frac{1}{E-E_j}=
\bar{g}(E)+\frac{1}{i\hbar}\sum_\gamma\sum_{n=1}^\infty
A_{n\gamma}\,\exp (inS_{\gamma}(E)/\hbar-i\pi n\nu_\gamma/2),
%
%
\label{gutz}
\end{equation}
where the amplitude $A_\gamma$ is defined as
\begin{equation}
A_{n\gamma}= \oint\,d\tau_\gamma |\det(M_{\gamma(\tau)}^n - I)|^{-1/2}.
\end{equation}
The term $\bar{g}(E)$ is a smooth function giving the mean density
of states. The double sum  runs over all distinct {\em periodic orbits}
in a {\em phase space}, labeled by $\gamma$, and over $n$,
the number of retracing each orbit.
The integer $\nu_\gamma$ is a phase shift :
in the case of finite systems with the Dirichlet boundary
conditions it counts the number of focal points and
twice the number of reflections off
the walls. $S_{\gamma}(E)$ is the action and the stability (monodromy)
matrix $M_{\gamma(\tau)}$ records the sensitivity
of the trajectory $\gamma$ at its given point $\gamma(\tau)$
to changes in initial conditions \cite{MG}.

The full power of the Gutzwiller approach has been
demonstrated in the discussion of the anisotropic Kepler
problem \cite{MG},  the scattering of a point particle from
three hard discs fixed on a plane - the so-called
 three disc repellor \cite{CRV},
hydrogen energy levels in a strong magnetic field
\cite{DD}, and
by the quantization of energy levels of the helium atom \cite{ERT}.
It is interesting to compare our diagrammatic rules with the
semiclassical Gutzwiller trace formula now. One sees immediately
that the number of periodic orbits
is {\em substantionally suppressed} in the exact expression. To visualize
(\ref{dfinal}) `semiclassically' each sphere $V_j$
containing a single scatterer is replaced by its center with regard
to which the single-site $t$ matrix is defined. Then it is natural that
only the {\em isolated} orbits (in the terminology of Ref. \cite{MB}) are
considered which connect {\em centers of different scatterers}.
{\em Nonisolated} orbits do not enter the exact expression,
for we have been interested in the calculation of
the {\em change} of the DOS and not of the DOS itself
as in  Refs. \cite{MB} and \cite{MG}. This might be a sign that
the convergence properties
of the Gutzwiller trace formula at the special case of
quantum billiards on a {\em torus} might be improved if
one calculates the semiclassical expansion
of the {\em change} of DOS directly. This does not concern
finite systems with the Dirichlet boundary conditions imposed that are
nonintegrable even without the presence of scatterers such as
the Bunimovitch stadium or quantum cavities \cite{BB}.


\section{Summary and conclusions}
To summarize, the photonic structure constants
$G^{jn}_{AL,A'L'}$ have been calculated and
basic MST and KKR equations for photons have been derived.
Our result for the structure constants
(\ref{structure}) is more symmetric
than that in  Ref. \cite{XZ} where, probably,
the trace was not kept whether the inward or outward formalism
was used.

A formal operator formalism for the Maxwell equation has been
presented.
The essential difference of the Maxwell equation (\ref{ME*})
with regard to the Schr\"{o}dinger equation is that,
in the former case, the potential from the point of view
of the Schr\"{o}dinger equation is {\em energy dependent}.
It was shown  [see (\ref{hami}) and (\ref{decom})] how to perform the
separation $\mbox{H}=
\mbox{H}_o+ \Gamma$ such that $\Gamma$ be {\em energy independent}.
The price one pays for the decomposition is that
the potential $\Gamma$ is itself a differential
operator multiplied by a spatially varying function
[see (\ref{decom})]. It reduces, however, to a multiplicative operator
on eigenfunctions of both $\mbox{H}$ and $\mbox{H}_o$.
The decomposition is necessary in order that the Green function
gives the density of states by the same formula as in the Schr\"{o}dinger
case. The Lippmann-Schwinger equations have been analyzed
within the operator formalism and properties of the Greens function have been
discussed.

The Lloyd and Smith on-the-energy-shell
multiple-scattering formalism \cite{PW} for the calculations
of the change $\triangle N(E)$ of the integrated density of states
induced by the presence of scatterers
has been analyzed in the spirit of the Gutzwiller approach
\cite{MG}.
The important message of our paper is that
one must not look  for the diagrams of Ref. \cite{PW} but for
{\em closed orbits in `phase space'} :
there are different orbits which give different contributions
to $\triangle N(E)$ but are described by the same diagram.
As a result, diagrammatic rules have been interpreted in
terms of `closed orbits'.

Our expression (\ref{dfinal}) shows that in the case of electromagnetic
waves the Krein-Friedel formula \cite{F} can be used as well.
Formula (\ref{dfinal}) gives $\triangle N(E)$ as the sum of
two contributions :  one that is determined solely in terms of
{\em single scattering properties}, and the second that is due
to {\em multiple-scattering effects}.
Therefore, in addition to the separation of purely geometric and purely
scattering properties as the standard KKR method does one can
separate the single-scattering and the multiple-scattering
contributions to the IDOS.
A comparison of the exact
expression (\ref{dfinal}) for $\triangle N(E)$ with the semiclassical
Gutzwiller trace formula (\ref{gutz}) (Ref. \cite{MG}) has been made.
The comparison shows that, in the special case of quantum billiards
on a torus, the number of closed orbits in the exact expression
to be  summed over is significantly reduced.

An  application of the above results to study the band structure of photons
and impurities in a photonic crystal will be given elsewhere \cite{AMo}.

\vspace*{0.8cm}

{\protect\Large\bf Acknowledgments}

\vspace*{0.3cm}

I should like to thank J. Kl\'{\i}ma, N. Pavloff, and
J. Petru for useful suggestions on the literature, and E. Hugues
and N. Pavloff for reading the manuscript and
discussions.
Partial support by the Internal Grant Agency of the Academy
of Sciences of the Czech Republic under Project No.
11086 and by the Grant Agency of the Czech Republic under
Project No. 202/93/0689 are gratefully acknowledged.

\appendix
\section{Appendix : Notations and definitions}
\label{nota}
Scalar spherical harmonics $Y_L$ are used as defined by Ref. \cite{J},
i.\,e., satisfying the {\em Condon-Shortley} convention
in which
\begin{equation}
Y^*_{lm} (\theta,\phi) =(-1)^m Y_{l,-m}(\theta,\phi),
\hspace*{0.8cm} Y_{lm}(-\theta,-\phi) = Y_{l,-m}(\theta,\phi).
\label{csc}
\end{equation}
Our definition of vector spherical harmonics coincides
up to factor $i$ with Ref.\ \cite{RN},
\begin{equation}
{\bf Y}^{(m)}_L = {\bf X}_L =\frac{1}{\sqrt{l(l+1)}}\,{\bf L}\,Y_L,
\end{equation}
\begin{equation}
{\bf Y}^{(e)}_L= ({\bf r}_o\times{\bf X}_L),
\end{equation}
\begin{equation}
{\bf Y}^{(o)}_L =iY_L{\bf r}_o,
\end{equation}
where ${\bf L}$ is the orbital angular momentum operator and
${\bf r}_o$ unit radius vector. They are all {\em normalized}
\begin{equation}
\langle {\bf Y}^{(a)}_L|{\bf Y}^{(a')}_{L'}\rangle =
\delta_{aa'}\delta_{LL'},
\end{equation}
and satisfy
\begin{equation}
\partial_r {\bf Y}^{(a)}_L =0.
\label{rader}
\end{equation}

One can show \cite{AMo} that
\begin{equation}
\mbox{\boldmath $\nabla$}\times{\bf X}_{L} =\frac{1}{r}{\bf Y}^{(e)}_L +
\frac{\sqrt{l(l+1)}}{r}{\bf Y}^{(o)}_L,
\end{equation}
\begin{equation}
\langle Y_{l'm'}|{\bf Y}^{(m)\alpha}_{lm}\rangle =
\frac{1}{\sqrt{l(l+1)}}\delta_{l'l} \delta_{m'm+\alpha}
T^\alpha_{lm},
\end{equation}
\begin{eqnarray}
\lefteqn{\langle Y_{l'm'}|{\bf Y}^{(e)\alpha}_{lm}\rangle =
-i\delta_{m'm+\alpha}\left\{
\sqrt{l+1}\,C^\alpha(l-1,m+\alpha,l,m)\delta_{l'l-1}\right.}
\hspace{4cm}\nonumber\\
&&
\left.\mbox{} +\sqrt{l}\,C^\alpha(l+1,m+\alpha,l,m)\delta_{l'l+1}\right\},
\end{eqnarray}
\begin{eqnarray}
\lefteqn{\langle Y_{l'm'}|{\bf Y}^{(o)\alpha}_{lm}\rangle =
-i\delta_{m'm+\alpha}\left\{
\sqrt{l} \,C^\alpha(l-1,m+\alpha,l,m)\delta_{l'l-1}\right.}
\hspace{3cm}\nonumber\\
&&
\left.\mbox{}-\sqrt{l+1}\,C^\alpha(l+1,m+\alpha,l,m)\delta_{l'l+1}\right\}.
\end{eqnarray}
Constants $T^\alpha_{lm}{}'^s$  are defined by the action of spherical
components of ${\bf L}^\alpha$ on spherical harmonics,
\begin{equation}
{\bf L}^\alpha Y_L = T^\alpha_{lm}Y_{lm+\alpha}.
\end{equation}
$C^\alpha(l'm'lm)$'s are defined by
\begin{equation}
\langle l'm'|V^\alpha|lm\rangle =
 C^\alpha(l'm'lm)\langle l'm'||V^\alpha||lm\rangle,
\end{equation}
where $\langle l'm'||V^\alpha||lm\rangle$ is the {\em reduced} matrix
element \cite{BL} that does not depend on $m$ and $m'$.
They can be expressed via $3j$ {\em symbols} \cite{BL},
\begin{equation}
C^\alpha(l'm'lm) = (-1)^{l'-m'}\left(
\begin{array}{rcc}
l'&1&l\\
-m'&\alpha&m
\end{array} \right).
\end{equation}
In all the above formulas $l\geq 1$ for there is neither electric nor
magnetic multipole with $l=0$.

The {\em Gaunt number} $C_{L_2L_1}^{L_3}$ is the matrix element \cite{KKR}
\begin{eqnarray}
\lefteqn{C_{L_2L_1}^{L_3}= \langle Y_{L_1}|Y_{L_2}|Y_{L_3}\rangle =
\oint Y^*_{L_1}Y_{L_2}Y_{L_3} do = }
\nonumber\\
&&
(-1)^{m_1}
\left[\frac{(2l_1+1)(2l_2+1)(2l_3+1)}{4\pi}\right]^{1/2}
\left(
\begin{array}{rcc}
l_1&l_2&l_3\\
-m_1&m_2&m_3
\end{array} \right)
\left(
\begin{array}{rcc}
l_1&l_2&l_3\\
0&0&0
\end{array} \right).
\label{gauntn}
\end{eqnarray}
In the Condon-Shortley
convention (\ref{csc}) that is adopted here they are all
{\em real} numbers. Due to the symmetry of $3j$ symbols one has
\begin{equation}
C_{L_2L_1}^{L_3}=C_{L_3L_1}^{L_2}=(-1)^{m_3}C_{L_1L_2}^{l_3,-m_3}=
C_{l_1,-m_1;l_2,-m_2}^{l_3,-m_3}.
\label{gauntprop}
\end{equation}
The first equality can be deduced
straightforwardly from the form of the integral in (\ref{gauntn}).
The second uses the fact that
\begin{equation}
\left(
\begin{array}{ccc}
j_1&j_2&j_3\\
0&0&0
\end{array} \right)=0
\end{equation}
whenever $J=j_1+j_2+j_3$ is {\em odd}. For $J$ {\em even} one has \cite{BL}
\begin{eqnarray}
\lefteqn{
\left(
\begin{array}{rcc}
j_1&j_2&j_3\\
0&0&0
\end{array} \right)=(-1)^{J/2}\left[
\frac{(J-2j_1)!(J-2j_2)!(J-2j_3)!}{(J+1)!}\right]^{1/2}
}\hspace*{3cm}\nonumber\\
&&
\frac{(J/2)!}{(J/2-j_1)!(J/2-j_2)!(J/2-j_3)!}\cdot
\label{3j000}
\end{eqnarray}
If $C_{L_2L_1}^{L_3}$ is  nonzero then
$(-1)^{l_1+l_2+l_3}=(-1)^{m_1+m_2+m_3}=1$.
Due to the properties (\ref{gauntprop}) of the Gaunt numbers and spherical
harmonics (\ref{csc}) one can write (\ref{stcon}) in several
equivalent forms,
\begin{eqnarray}
\lefteqn{g_{L'L}({\bf R})
 = - (-1)^l i^{l+l'+1}  4\pi  \sigma
\sum_{L_1} C_{L'L}^{L_1} i^{l_1}
 h^+_{l_1}(\sigma R) Y^*_{L_1}({\bf R})
}
\nonumber\\
&&
 = -(-1)^l i^{l+l'+1}  4\pi \sigma\sum_{L_1} C_{LL'}^{L_1} i^{l_1}
 h^+_{l_1}(\sigma R) Y_{L_1}({\bf R})
\nonumber\\
&&
 =- (-1)^{l'}i^{l+l'+1}  4\pi \sigma\sum_{L_1} C_{LL'}^{L_1}  i^{l_1}
 h^+_{l_1}(\sigma R) Y_{L_1}(-{\bf R}).
\label{stcon1}
\end{eqnarray}

\section{Appendix : Basic MST identities}
\label{basic}
\begin{equation}
W(j_l,h^+_l)=iW\left[j_l(\sigma r),n_l(\sigma r)\right]
=\frac{i}{\sigma r^2},
\label{wro}
\end{equation}
\begin{equation}
\langle {\bf J}^\alpha_{AL}|{\bf J}^\alpha_{A'L'}\rangle=
\delta_{AA'}\delta_{LL'}N_A^l=\delta_{AA'}\delta_{LL'}\left\{
\begin{array}{cc}
j_l^2&A=M\\
\frac{1}{(2l+1)}\left[(l+1)j_{l-1}^2+lj_{l+1}^2\right] &A=E,
\end{array}
\right.
\end{equation}
\begin{equation}
\langle {\bf J}^{\alpha}_{ML} |Y_{L'} \rangle =
j_l\,\frac{1}{\sqrt{l(l+1)}}\,T^\alpha_{lm}
\delta_{l'l}\delta_{m'm+\alpha},
\label{causetl}
\end{equation}
\begin{eqnarray}
\lefteqn{\langle {\bf J}^{\alpha}_{EL} |Y_{L'} \rangle =
 \delta_{m'm+\alpha}\left\{\sqrt{l+1}\,
j_{l-1}C^\alpha(l-1,m+\alpha,l,m)
\delta_{l'l-1}\right.}\hspace{6cm}\nonumber\\
&&
\left.\mbox{}- \sqrt{l}\,j_{l+1}C^\alpha(l+1,m+\alpha,l,m)
\delta_{l'l+1}\right\}
\nonumber
\end{eqnarray}
\begin{equation}
=
 \delta_{m'm+\alpha}\left\{
j_{l-1}\bar{C}^\alpha(l,-1,m)
\delta_{l'l-1}
+ j_{l+1}\bar{C}^\alpha(l,1,m)
\delta_{l'l+1}\right\}.
\label{causecl}
\end{equation}
All the above formulas follow directly from the definitions of
magnetic (\ref{mmulti}) and electric (\ref{elmd}) multipoles and scalar
product formulas of the preceding appendix.
Now, one can show that
\begin{equation}
\langle Y_{L'}|j_{l'}\partial_r - j_{l'}'|{\bf J}^\alpha_{AL}\rangle
=\langle Y_{L'}|W(j_{l'},{\bf J}^\alpha_{AL})\rangle
=
\langle Y_{L'}|W(h^+_{l'},{\bf H}^+_{AL})\rangle=0.
\end{equation}
For magnetic multipoles the result is
zero for $l\neq l'$ as the consequence of scalar product formulas,
and zero for $l'=l$ for it is proportional to the Wronskian
of identical functions. Similarly, for electric multipoles
one needs to check only the case of $l'=l\pm 1$ since otherwise
the result is, as above,  zero due to scalar product formulas.
In the case of $l'=l+1$, for example,
one finds in principle terms that are either proportional to
$W(j_{l+1}, j_{l+1})$ or $W(j_{l+1},j_{l-1})$.
The former then vanish because of vanishing Wronskian,
the latter because of vanishing of
the numerical prefactor in front of it.
As a direct consequence of the above considerations one has
\begin{equation}
\langle Y_{L'}|W(h^+_{l'},{\bf J}_{AL})\rangle=
-\langle Y_{L'}|W(j_{l'},{\bf H}^+_{AL})\rangle
=- \langle Y_{L'}|j_{l'}\partial_r - j_{l'}'|{\bf H}^+_{ML}\rangle,
\label{chan}
\end{equation}
\begin{equation}
\langle Y_{L'}|W(j_{l'},{\bf H}^{+\alpha}_{ML})\rangle=
\frac{i}{\sigma r^2}\frac{1}{\sqrt{l(l+1)}}
T^\alpha_{lm}\delta_{l'l}\delta_{m'm+\alpha},
\end{equation}
\begin{eqnarray}
\langle Y_{L'}|W(j_{l'},{\bf H}^{+\alpha}_{EL})\rangle
=\frac{i}{\sigma r^2}\delta_{m'm+\alpha}\left[\sqrt{l+1}\,
C^\alpha(l-1,m+\alpha,l,m)\delta_{l'l-1}
\right.\nonumber\\
\left.\mbox{} -\sqrt{l}\, C^\alpha(l+1,m+\alpha,l,m)\delta_{l'l+1}\right]
\nonumber\\
=\frac{i}{\sigma r^2}\delta_{m'm+\alpha}\left[
\bar{C}^\alpha(l,-1,m)\delta_{l'l-1}
+\bar{C}^\alpha(l,1,m)\delta_{l'l+1}\right].
\end{eqnarray}

Ignoring for a while  $j_l$ terms and
multiplicative factors one finds
by using the above formulas that
$\langle Y_{L'}|W(j_{l'},{\bf H}^+_{AL})\rangle
=\langle Y_{L'}| {\bf J}_{AL}\rangle$.
This indicates that, by using (\ref{chan}),
\begin{equation}
\sum_{L"}\langle {\bf J}_{AL}|Y_{L"}\rangle\cdot
\langle Y_{L"}|W(h^+_{l"},{\bf J}_{A'L'})\rangle=
\sum_{L"}\langle {\bf J}_{AL}|Y_{L"}\rangle\cdot
\langle Y_{L"}|{\bf J}_{A'L'}\rangle=
\langle {\bf J}_{AL}|{\bf J}_{A'L'}\rangle.
\end{equation}
Including all the factor one indeed confirms that
\begin{eqnarray}
\sum_{L"}j_{l"} \langle {\bf J}_{AL}|Y_{L"}\rangle\cdot
\langle Y_{L"}|W(h^+_{l"},{\bf J}_{A'L'})\rangle=
-\frac{i}{\sigma r^2}\delta_{AA'}\delta_{LL'}N_A^l.
\end{eqnarray}
The result is a consequence of two identities,
\begin{equation}
\sum_{\alpha=-1}^1 \bar{C}^{\alpha 2}(l,1,m)=\frac{l+1}{2l+1}
\hspace*{0.8cm}\mbox{and}\hspace*{0.8cm}
\sum_{\alpha=-1}^1 \bar{C}^{\alpha 2}(l,-1,m)=\frac{l}{2l+1}\cdot
\end{equation}
They immediately imply
\begin{equation}
\sum_{\stackrel{p=-1}{\scriptscriptstyle p\neq 0}}^1
\sum_{\alpha=-1}^1 \bar{C}^{\alpha}(l,p,m){}^2 =1.
\label{rel1}
\end{equation}
Similarly, one can find that
\begin{equation}
\sum_{\alpha =-1}^1 \bar{T}^{\alpha}_{lm}{}^2 =1.
\label{rel2}
\end{equation}
The last two identities then prove the ``unitarity'' of
${\protect\bf\tau}$ and  ${\bf c}$ matrices.


\begin{thebibliography}{99}
\bibitem{Ham}Despite suggestive notation $\mbox{H}$
is not the Hamiltonian in the dynamical sense
since the nonstationary predecessor of (\ref{ME*}) involves
the second order  derivative in time. This must be keep in mind
when one wants to represent the resolvent by path integral.
\end{thebibliography}
\end{document}